\documentclass[journal]{IEEEtran}
\usepackage[numbers]{natbib}
\usepackage[pdftex]{graphicx}
\graphicspath{{./figs}}
\usepackage{amsmath}
\usepackage{algorithmicx}
\usepackage{array}
\usepackage{stfloats}
\usepackage{url}
\usepackage{hyperref}
\usepackage{hypernat}

\begin{document}

\title{Joint Track Machine Learning: An Autonomous Method of Measuring TKA Kinematics from Single-Plane Images}%

\author{Andrew Jensen, Paris Flood, Lindsey Palm-Vlasak, Will Burton, Paul Rullkoetter, Scott Banks%
\thanks{This work is generously supported by the McJunkin Family Charitable Foundation}%
\thanks{A. Jensen, L. Palm-Vlasak, and S. Banks are from the Department of Mechanical and Aerospace Engineering at the University of Florida}%
\thanks{P. Flood is from the Department of Computer Science at the University of Cambridge}%
\thanks{W. Burton and P. Rullkoetter are from the School of Engineering and Computer Science at the University of Denver}}%

\markboth{A Preprint}{Joint Track Machine Learning}

\maketitle

\begin{abstract}
    Dynamic radiographic measurement of 3D TKA kinematics has provided important information for implant design and surgical technique for over 30 years. However, current methods of measuring TKA kinematics are too cumbersome or time-consuming for practical clinical application. Even state-of-the-art techniques require human-supervised initialization or human supervision throughout the entire optimization process. Elimination of human supervision could potentially bring this technology into clinical practicality. Therefore, we propose a fully autonomous pipeline for quantifying TKA kinematics from single-plane imaging. First, a convolutional neural network segments the femoral and tibial implants from the image. Second, segmented images are compared to Normalized Fourier Descriptor shape libraries for initial pose estimates. Lastly, a Lipschitzian optimization routine minimizes the difference between the segmented image and the projected implant. This technique reliably reproduces human-supervised kinematics measurements from internal datasets and external validation studies, with RMS differences of less than 0.7mm and 4° for internal studies and 0.8mm and 1.7° for external validation studies. This performance indicates that it will soon be practical to perform these measurements in a clinical setting.
\end{abstract}


\section{Introduction}
Total Knee Arthroplasty (TKA) is a standard procedure for alleviating symptoms related to osteoarthritis in the knee. In 2018, orthopaedic surgeons performed more than 715,000 TKA operations in the United States \cite{agencyforhealthcareresearchandqualityHCUPFastStats}. This number is projected to increase to 3.48 million by 2030 \cite{kurtzProjectionsPrimaryRevision2007} due to an aging population and increased obesity rates. While TKA largely relieves symptomatic osteoarthritis, roughly 20\% of TKA patients express postoperative dissatisfaction, citing mechanical limitations, pain, and instability as the leading causes \cite{bakerRolePainFunction2007,bournePatientSatisfactionTotal2010,scottPredictingDissatisfactionFollowing2010}. Standard methods of musculoskeletal diagnosis cannot quantify the dynamic state of the joint, either pre- or post-operatively; clinicians must rely on static imaging (radiography, MRI, CT) or qualitative mechanical tests to determine the condition of the affected joint, and these tests cannot easily be performed during weight-bearing or dynamic movement when most pain symptoms occur. Unfortunately, most of the tools used to quantify 3D dynamic motion are substantially affected by soft-tissue artifacts \cite{gaoInvestigationSoftTissue2008,stagniQuantificationSoftTissue2005,linEffectsSoftTissue2016}, are prohibitively time-consuming or expensive \cite{daemsValidationThreedimensionalTotal2016a}, or cannot be performed with equipment available at most hospitals.

Model-image registration is a process where a 3D model is aligned to match an object’s projection in an image \cite{brownSurveyImageRegistration1992}. Researchers have performed model-image registration using single-plane fluoroscopic or flat-panel imaging since the 1990s. Early methods used pre-computed distance maps \cite{lavalleeRecoveringPositionOrientation1995,zuffiModelbasedMethodReconstruction1999}, or shape libraries \cite{banksAccurateMeasurementThreedimensional1996,wallaceAnalysisThreedimensionalMovement1980,wallaceEfficientThreedimensionalAircraft1980} to match the projection of a 3D implant model to its projection in a radiographic image. With increasing computational capabilities, methods that iteratively compared implant projections to images were possible \cite{mahfouzRobustMethodRegistration2003,floodAutomatedRegistration3D2018,loweFittingParameterizedThreedimensional1991}. Most model-image registration methods provide sufficient accuracy for clinical joint assessment applications, including natural and replaced knees \cite{banksVivoKinematicsCruciateretaining1997,banks2003HapPaul2004,komistekVivoFluoroscopicAnalysis2003,burtonAutomaticTrackingHealthy2021}, natural and replaced shoulders \cite{kijimaVivo3dimensionalAnalysis2015,mahfouzVivoDeterminationDynamics,matsukiVivo3dimensionalAnalysis2011,sugiComparingVivoThreedimensional2021}, and extremities \cite{cenniKinematicsThreeComponents2012,cenniFunctionalPerformanceTotal2013,deaslaSixDOFVivo2006}. One of the main benefits of this single-plane approach is that suitable images can be acquired with equipment found in most hospitals. The main impediment to implementing this approach into a standard clinical workflow is the time and expense of human operators to supervise the model-image registration process. These methods require either (1) an initial pose estimate \cite{floodAutomatedRegistration3D2018,loweFittingParameterizedThreedimensional1991}, (2) a pre-segmented contour of the implant in the image \cite{brownSurveyImageRegistration1992,lavalleeRecoveringPositionOrientation1995}, or (3) a human operator to assist the optimization routine out of local minima \cite{mahfouzRobustMethodRegistration2003}. Each of these requirements makes model-image registration methods impractical for clinical use. Even state-of-the-art model-image registration techniques \cite{floodAutomatedRegistration3D2018} require human initialization or segmentation to perform adequately.

\begin{figure*}[ht]
    \centering
    \includegraphics[width = \textwidth]{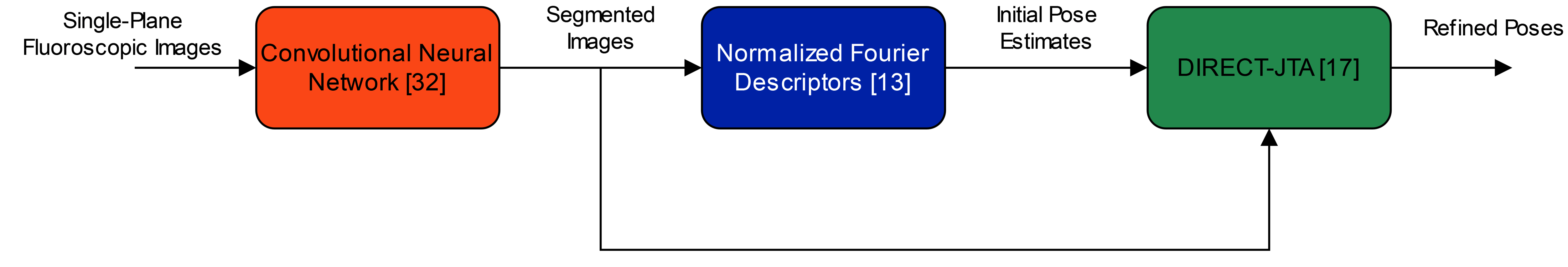}
    \caption{An overview of the pipeline for autonomous measurements of total knee arthroplasty kinematics. First, the data is processed through a convolutional neural network to locate the pixels belonging to the femoral and tibial implants \cite{wangDeepHighResolutionRepresentation2020}, then, Normalized Fourier Descriptor shape libraries are used to determine and initial pose estimate \cite{banksAccurateMeasurementThreedimensional1996}, and lastly, DIRECT-JTA \cite{floodAutomatedRegistration3D2018} is run on those segmented images using the NFD estimates as initializations for pose.}
    \label{fig:pipeline}
\end{figure*}

Machine learning algorithms automate the process of analytical model building, utilizing specific algorithms to fit a series of inputs to their respective outputs. Neural networks are a subset of machine learning algorithms that utilize artificial neurons inspired by the human brain’s connections \cite{marrEarlyProcessingVisual1976}. These networks have shown a great deal of success in many computer vision tasks, such as segmentation \cite{chanHistoSegNetSemanticSegmentation2019,wangDeepHighResolutionRepresentation2020,ronnebergerUNetConvolutionalNetworks2015}, pose estimation \cite{wuDeepGraphPose2020,kendallGeometricLossFunctions2017}, and classification \cite{krizhevskyImageNetClassificationDeep2017,qiPointNetDeepHierarchical2017,qiPointNetDeepLearning2017}. These capabilities might remove the need for human supervision from TKA model-image registration. Therefore, we propose a three-stage data analysis pipeline (Fig. \ref{fig:pipeline}) where a convolutional neural network (CNN) is used to segment, or identify, the pixels belonging to either a femoral or tibial component. Then, an initial pose estimate is generated comparing the segmented implant contour to a pre-computed shape library. Lastly, the initial pose estimate serves as the starting point for a Lipschitzian optimizer that aligns the contours of a 3D implant model to the contour of the CNN-segmented image.

This paper seeks to answer the following three questions: (1) How well does a convolutional neural network segment the femoral and tibial implants from fluoroscopic and flat-panel images? (2) Can a Fourier descriptor-based pose estimation method produce useful initial guesses of 3D implant pose from the CNN-segmented images? (3) Can the Lipschitzian optimizer, given reasonable initial guesses, replicate human-supervised TKA kinematic measurements? 

\section{Methods}

\begin{figure}[h]
    \centering
    \includegraphics[width = \linewidth]{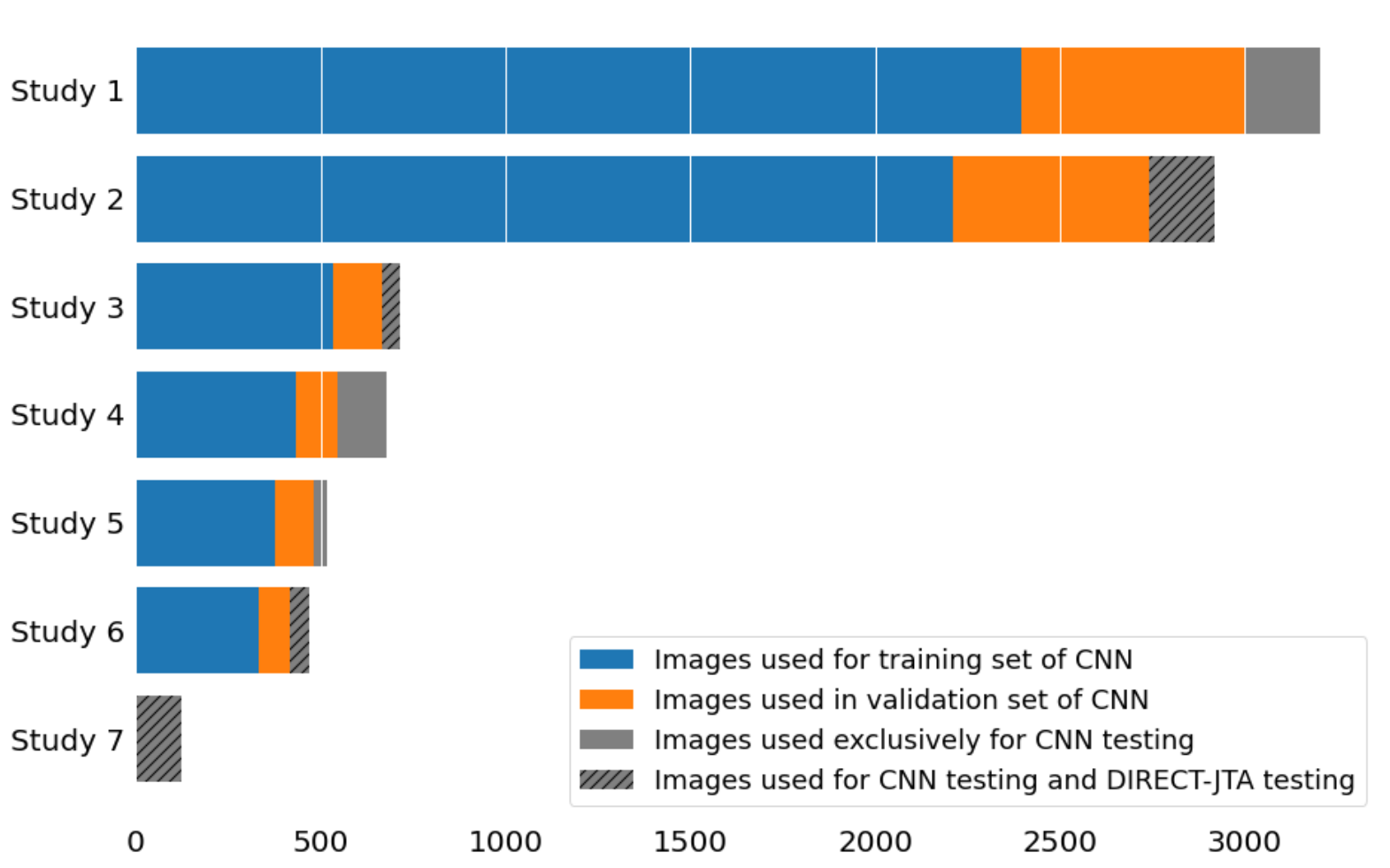}
    \caption{Data from seven studies were used to train and test the TKA kinematics measurement pipeline. Color coding in the figure identifies how many images were used for the training, validation, and testing functions. Images from the seventh study were used exclusively for testing the measurement pipeline that was trained using images from the other six studies.}
    \label{fig:dataset}
\end{figure}

Data from seven previously reported TKA kinematics studies were used for this study \cite{kefalaAssessmentKneeKinematics2017,palm-vlasakMinimalVariationTop2022,okamotoVivoKneeKinematics2011,watanabeKneeKinematicsAnterior2013a,jennyREGISTRATIONKNEEKINEMATICS2015,watanabeInvivoKinematicsHighflex2016,scottCanTotalKnee2016}. These studies utilized single-plane fluoroscopy or flat-panel imaging to measure tibiofemoral implant kinematics during lunge, squat, kneel, and stair climbing movements from 8248 images in 71 patients with implants from 7 manufacturers, including 36 distinct implants. From each of these studies, the following information was collected: (1) deidentified radiographic images, (2) x-ray calibration files, (3) manufacturer-supplied tibial and femoral implant surface geometry files (STL format), and (4) human supervised kinematics for the tibial and femoral components in each of the images. CNNs were trained with images from six of the studies using a transfer-learning paradigm with an open-source network \cite{wangDeepHighResolutionRepresentation2020}. CNN performance was tested using two image collections: a standard test set including images from the six studies used for training and a wholly naïve test set using images from the seventh study, where the imaging equipment and implants were different from anything used in training (Fig. \ref{fig:dataset}). We used both test image sets to compare human-supervised kinematics with autonomously measured kinematics. Separately, two independent groups utilized our software to assess the accuracy of TKA kinematics measurements compared to their previously reported reference standard systems using RSA \cite{teeterQuantificationVivoImplant2013} or motion capture \cite{daemsValidationThreedimensionalTotal2016a}.

\subsection{Image Segmentation}
\begin{figure*}[h]
    \centering
    \includegraphics[width = 0.75\textwidth]{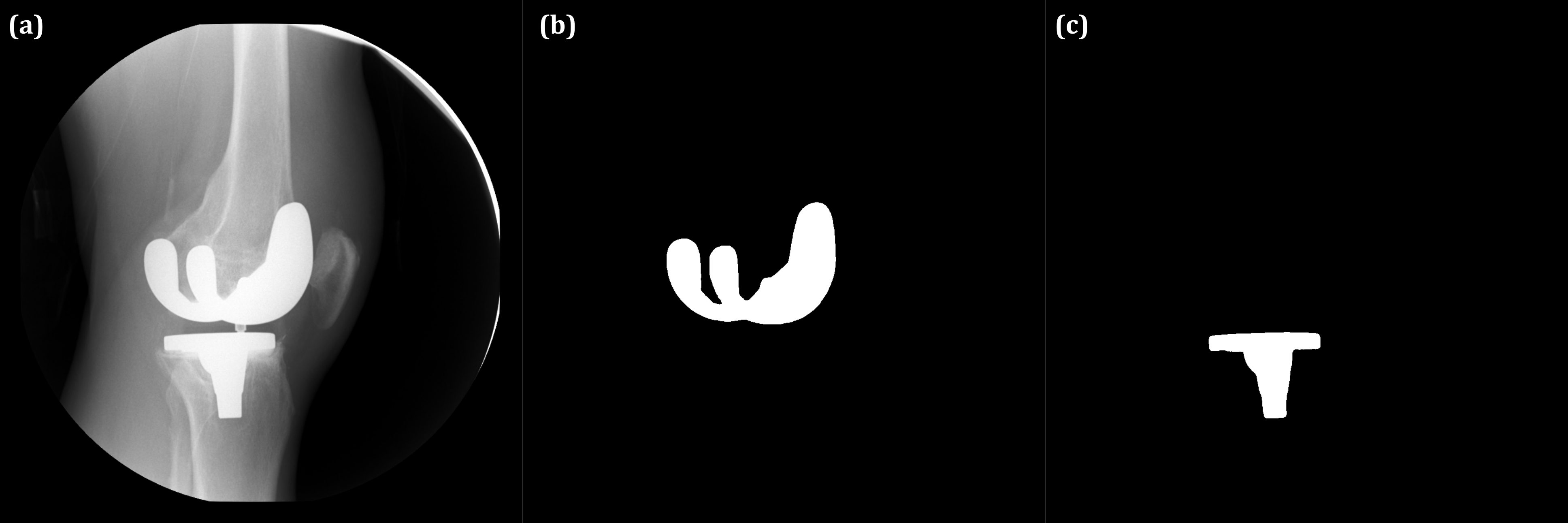}
    \caption{A representative fluoroscopic images is shown (a) with corresponding femoral (b) and tibial (c) ground-truth images created by flat-shaded projections of registered implant models.}
    \label{fig:seg-labels}
\end{figure*}

Images were resized and padded to 1024x1024 pixels. Images containing bilateral implants had the contralateral knee cropped from the image. Segmentation labels were created by taking the human-supervised kinematics for each implant and generating a flat-shaded ground-truth projection image (Fig. \ref{fig:seg-labels}). Two neural networks \cite{wangDeepHighResolutionRepresentation2020} were trained to segment the tibial and femoral implants, respectively, from the x-ray images. Each network was trained using a random 6284/1572 (80/20) training/validation split. Augmentations were introduced in the training pipeline to improve the network's generalization to new implants and implant types \cite{buslaevAlbumentationsFastFlexible2020}. Each neural network was trained on an NVIDIA A100 GPU for 30 epochs. The performance of the segmentation networks was measured using the Jaccard Index \cite{jaccardDISTRIBUTIONFLORAALPINE1912}. This calculates the intersection between the estimated and ground-truth pixels over the union of both sets of pixels. The ideal Jaccard index is 1.

\subsection{Initial Pose Estimates}

\begin{figure*}[ht]
    \centering
    \includegraphics[width = \linewidth]{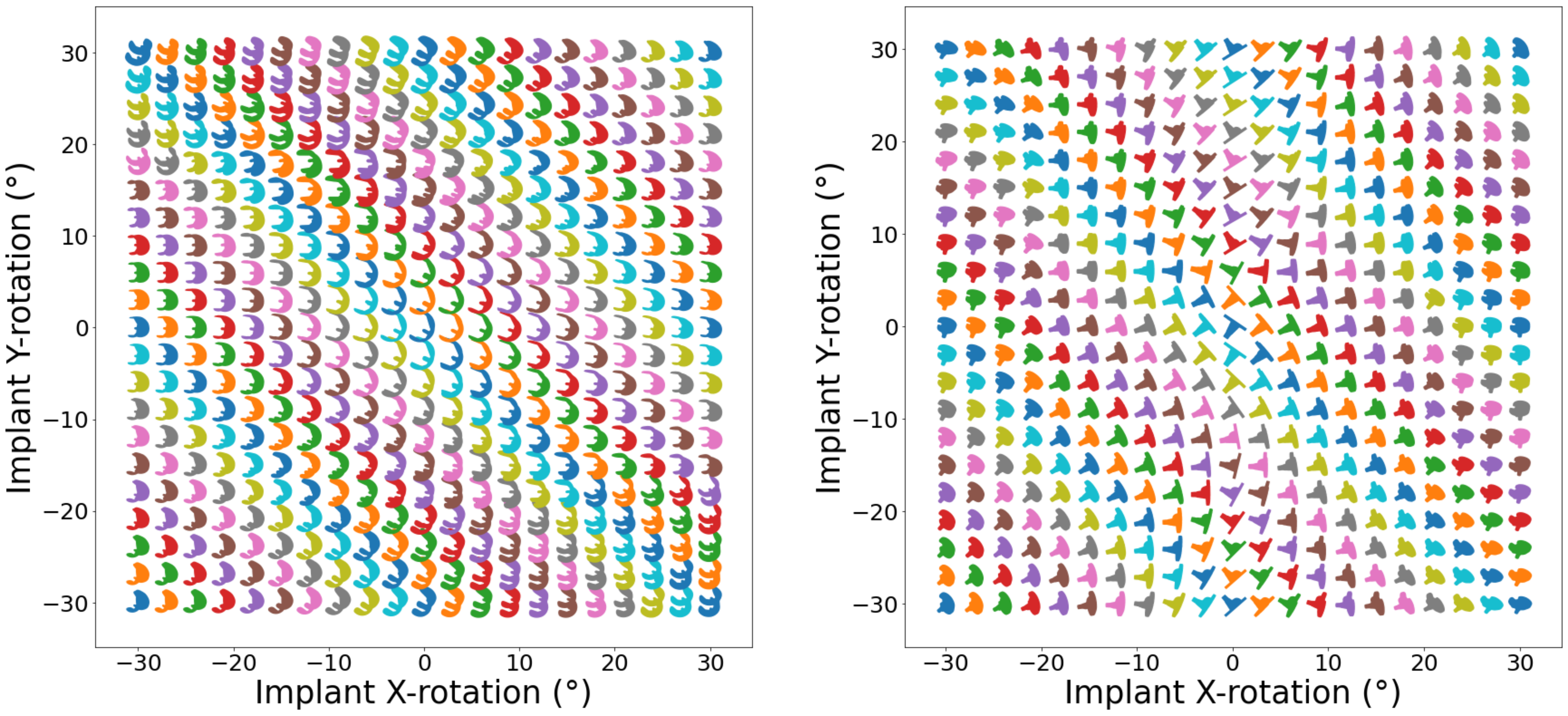}
    \caption{Femoral (left) and tibial (right) NFD shape libraries were generated to capture the variation in projection silhouette geometry with out-of-plane rotation \cite{banksAccurateMeasurementThreedimensional1996}. Initial pose estimates were generated by comparing the NFD contour from the x-ray image to the shape library.}
    \label{fig:nfd-lib}
\end{figure*}

Initial pose estimates were generated from bounding contours of the CNN-segmented implant regions using Normalized Fourier Descriptor (NFD) shape libraries \cite{banksAccurateMeasurementThreedimensional1996,wallaceAnalysisThreedimensionalMovement1980,wallaceEfficientThreedimensionalAircraft1980}. Shape libraries were created by projecting 3D implant models using the corresponding x-ray calibration parameters with ±30° ranges for the out-of-plane rotations at 3° increments (Fig. \ref{fig:nfd-lib}). Pose estimates were determined as previously described \cite{banksAccurateMeasurementThreedimensional1996} NFD-derived femoral and tibial implant poses were transformed to anatomic joint angles and translations \cite{groodJointCoordinateSystem1983} and compared to the human-supervised kinematics for the same images using RMS differences for each joint pose parameter. The performance of this method was also assessed using flat-shaded projection images with perfect segmentation as a ground-truth reference standard.

\subsection{Pose Refinement}
A modified Dividing Rectangles (DIRECT) algorithm called DIRECT-JTA  \cite{floodAutomatedRegistration3D2018} generated the final pose estimates. This method of Lipschitzian optimization divides the search into three stages, the “trunk,” “branch,” and “leaf.” Each of the three stages was assigned distinct cost function parameters and search regions. The cost function used a computationally efficient L1-norm between the dilated contour from the segmentation label and the projected implant. Successively decreasing the dilation coefficient allowed the optimization routine to escape local minima, and the leaf branch served to find the optimal out-of-plane translation. Transversely symmetric tibial implants posed problems during registration because two distinct poses produced roughly identical projections \cite{kendallShapeManifoldsProcrustean1984}. Because of this pose ambiguity, the tibial implant was always optimized after the non-symmetric femoral implant. In addition to the dilation metric, the tibial mediolateral translation and varus/valgus rotations relative to the femur were penalized. Final implant poses were transformed into knee joint rotations and translations \cite{groodJointCoordinateSystem1983} and compared to the human-supervised kinematics for the same images using RMS differences for each joint pose parameter. Squared differences between data sets were compared using one-way MANOVA with post-hoc multiple pair-wise comparisons using the Games-Howell test (R v4.2.0 using R Studio, rstatix, and stats).

\subsection{Pose Ambiguities and Registration Blunders}
A blunder was defined as an image frame with the squared sum of rotation differences greater than 5° between autonomous and human-supervised measures. These blunder frames contain errors considerably larger than would be clinically acceptable and warrant further exploration. Blunders were analyzed with respect to the tibial implant’s apparent varus/valgus rotation relative to the viewing ray (Fig. \ref{fig:histo-pdf}). A probability density function and cumulative density function were calculated for the blunder likelihood. Due to the high likelihood of blunders in this region, an ambiguous zone was defined for all apparent tibial varus/valgus-rotation less than 3.6 degrees, which is the mean + 1std of the blunder distribution (Fig. \ref{fig:histo-pdf}). Squared measurement differences between images inside and outside the ambiguous zone were also compared using one-way MANOVA with post-hoc multiple pair-wise comparisons using the Games-Howell test.

\begin{figure*}[hb]
    \centering
    \includegraphics[width = \linewidth]{figs/fig5-histo.png}
    \caption{The histogram (left) shows the correctly registered frames (Hits, blue) and incorrectly registered frames  (Blunders, orange) plotted as a function of the apparent tibial varus/valgus angle relative to the viewing raw. The probability plot (right) shows the distribution of blunders (solid orange) and the cumulative probability of blunders (dotted orange). The Ambiguous Zone is defined as apparent tibial varus/valgus rotations less than the mean + one standard deviation of the blunder probability distribution, capturing approximately 85 \% of the blunders.}
    \label{fig:histo-pdf}
\end{figure*}

\section{Results}
CNN segmentation of standard test set images produced Jaccard indices of 0.936 for the femoral and 0.883 for the tibial components. CNN segmentation performance on the completely naïve test set was lower, 0.715 and 0.753, respectively.

The initial pose estimates were within the range of convergence for the DIRECT-JTA optimizer and offered a robust initialization for optimization (Table 1). The RMS differences for initial pose estimates on ground-truth images were smaller (better) than for CNN-segmented images, but the differences were mostly within a few millimeters or degrees. Due to poor sensitivity for measuring out-of-plane translation with monocular vision, the mediolateral translation had the largest RMS differences for both image types.

\begin{figure*}[ht]
    \centering
    \includegraphics[width = 0.85\textwidth]{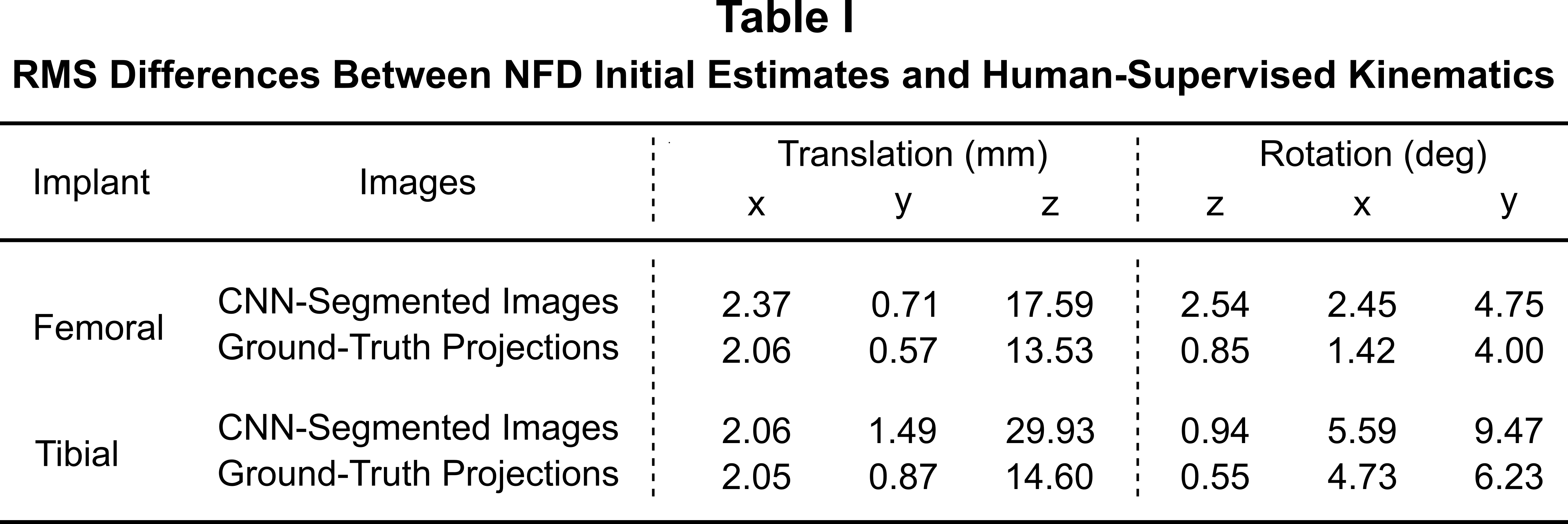}
\end{figure*}

RMS differences between DIRECT-JTA optimized kinematics and human-supervised kinematics were sub-millimeters for all in-plane translations (Table II). Mediolateral translations and out-of-plane rotation differences were smaller when the pose of the tibia was outside the ambiguous zone. The RMS differences for the completely naïve test set were within 0.5 mm or 0.5 deg compared to the standard test set, indicating similar performance on the entirely novel dataset.

\begin{figure*}[hb]
    \centering
    \includegraphics[width = 0.93\textwidth]{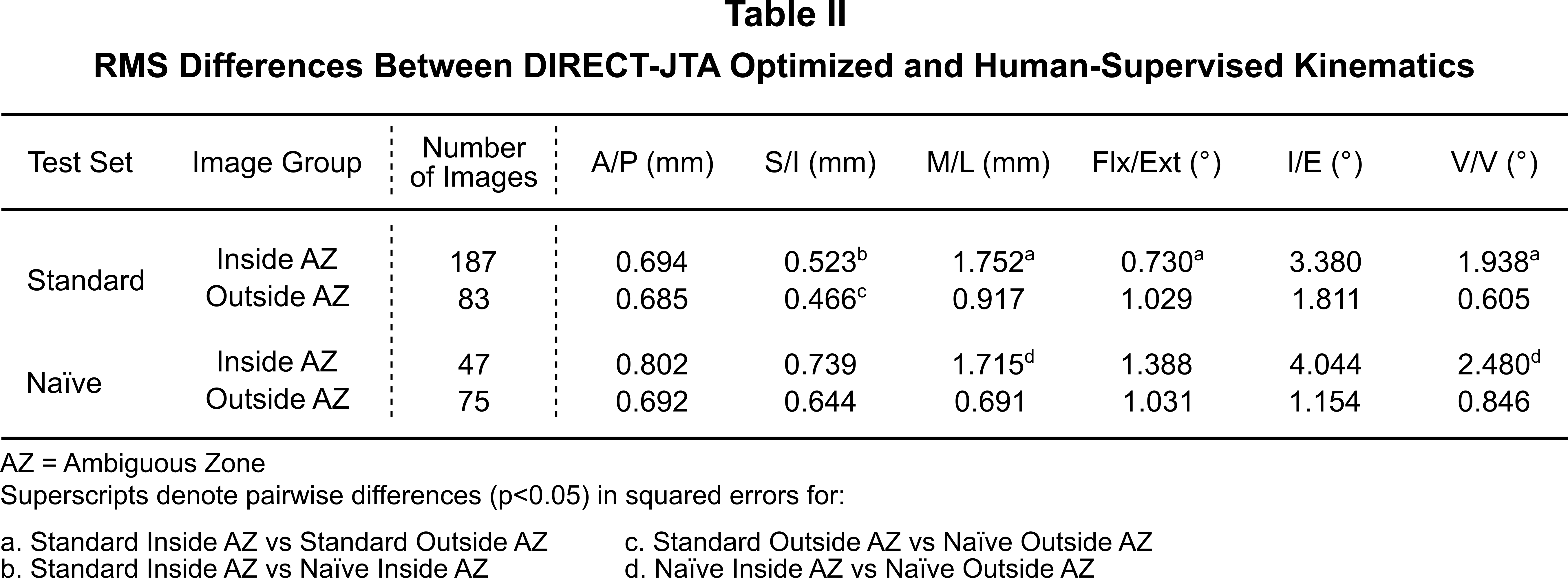}
    \label{tab:direct-rms}
\end{figure*}

\begin{figure}[!h]
    \centering
    \includegraphics[width = \linewidth]{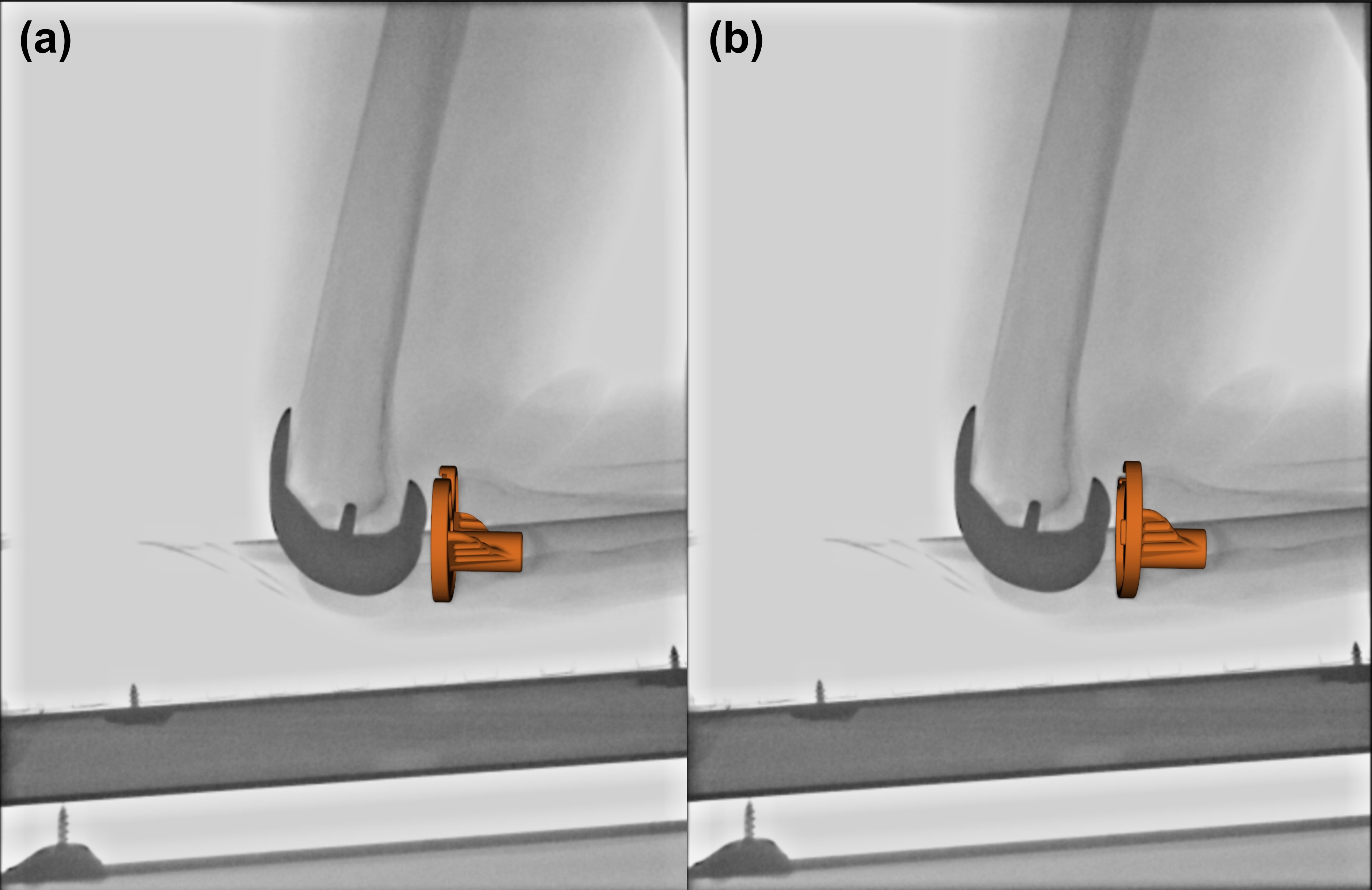}
    \caption{The figure shows the same radiographic image with two registered tibial implant poses: (a) shows a correctly registered tibial implant, while (b) shows an implant caught in a local cost function minimum corresponding to a nearly symmetric pose.}
    \label{fig:sym-trap}
\end{figure}

There was one femoral blunder and 43 tibial blunders out of 392 test images. Using the definition of the ambiguous zone as apparent tibial varus/valgus rotation less than 3.6 deg, 11\% of images have a tibial blunder within this zone, compared to 3.2\% outside. Sixty-six percent of tibial blunders were due to symmetry ambiguities (Fig \ref{fig:sym-trap}).

One-hundred thirteen image pairs from an RSA study of TKA were used to independently assess the accuracy of the autonomous kinematics measurement for single-plane lateral TKA images. RMS errors were 0.8mm for AP translation, 0.5mm for SI translation, 2.6mm for ML translation, 1.0° for flexion-extension, 1.2° for abduction-adduction, and 1.7° for internal-external rotation. At a different institution, 45 single-plane radiographic images were acquired with an instrumented sawbones phantom that was independently tracked using motion capture. Comparing the motion capture and autonomously measured radiographic kinematics, the RMS errors were 0.72mm for AP translation, 0.31mm for SI translation, 1.82mm for ML translation, 0.56° for flexion-extension, 0.63° for abduction-adduction, and 0.84° for internal-external rotation.

\section{Discussion}
Dynamic radiographic measurement of 3D TKA kinematics has provided important information for implant design and surgical technique for over 30 years. Many surgeons have expressed an interest in utilizing this type of measurement in their clinical practices; however, current methods are impractical. We developed a completely autonomous TKA kinematics measurement pipeline that can potentially provide a practical method for clinical implementation. This study sought to answer three questions, (1) How well does a neural network segment TKA implants from fluoroscopic and flat-panel images? (2) How well can an NFD shape library estimate the pose of a TKA implant given a CNN-segmented image? And (3) How well does a Lipschitzian optimization routine replicate human-supervised kinematics for TKA implants given an approximate initial guess? 

CNN image segmentation of TKA implants worked well, with Jaccard indices greater than 0.88 for the standard test set, and greater than 0.71 for the naïve test set. Segmentation performance for the standard test set outperformed published examples by 0.05-0.1 Jaccard points \cite{zhouUNetNestedUNet2018,rodriguesDeepSegmentationLeverages2019}, with the naïve test set on par with other segmentation examples. The most notable decrease in segmentation performance occurred along the perimeter of the segmented pixel region, especially in areas where implant projections occluded each other. These imperfectly segmented perimeter regions likely affect the initial pose estimate and the DIRECT-JTA optimization solution since both methods rely heavily on the segmented implant boundary. Further improvements can be made for the perimeter segmentation results by introducing intelligent augmentations during training using generative models \cite{hatayaFasterAutoAugmentLearning2019} and performing neural network bolstered contour improvement strategies \cite{yuanSegFixModelAgnosticBoundary2020}. 

Our initial pose estimates were satisfactory as an initialization for the DIRECT-JTA optimization, falling within the convergence region of ±30° \cite{floodAutomatedRegistration3D2018}. However, the performance for the ground-truth projections was not as good as the cited method \cite{banksAccurateMeasurementThreedimensional1996}, which achieved errors of less than 1mm for in-plane translation and 2° for rotation. The cited method utilized an additional refinement step for the NFD estimation, interpolating the apparent out-of-plane angles between nearest shapes in the library. This extra step was not done because only approximate initial pose estimates were needed. In addition, the current study incorporated a vastly larger set of implant shapes (36 vs. 2) and image quality and calibration variations. Distinct implant shapes manifest unique normalization maps, where there can be discontinuities or jumps in normalization angles which affect the best-fitting library entry (Fig. \ref{fig:nfd-lib}) \cite{wallaceAnalysisThreedimensionalMovement1980,wallaceEfficientThreedimensionalAircraft1980}. These details are easily upgraded with additional code using previously reported methods but were not pursued because the initial pose results were well within the DIRECT-JTA convergence region. The initial pose estimates for the CNN-segmented images were not as good as for the ground-truth projections. This follows directly from the fact that the perimeter of the segmented implants was not as accurately rendered, leading to poorer results with the edge-based NFD method. Finally, the out-of-plane translation estimates were relatively poor for both ground-truth projects and CNN-segmented images. This translation estimate is extremely sensitive to model projection and edge detection details and can be adjusted for better results if required. 

RMS differences between human-supervised and DIRECT-JTA optimized kinematics demonstrate the two methods provide similar results. In-plane translation differences of less than 0.8mm and out-of-plane less than 1.8 mm, indicate good consistency in determining the relative locations of TKA implants. Rotation differences of 4° or less for frames within the ambiguous zone, and less than 1.7° for frames outside the ambiguous zone, indicate joint rotation measures with sufficient resolution to be clinically useful. We observed two important characteristics in the measurement comparisons that will affect future implementations and use. First, we identified an ambiguous zone of apparent tibial rotations wherein there is a higher incidence of registration errors. These errors resulted in significant differences in measurement performance for the out-of-plane translations and rotations. This phenomenon, resulting from the nearly symmetric nature of most tibial implants \cite{lavalleeRecoveringPositionOrientation1995,zuffiModelbasedMethodReconstruction1999,banksAccurateMeasurementThreedimensional1996,mahfouzRobustMethodRegistration2003,floodAutomatedRegistration3D2018} prompts either practical modification to imaging protocols to bias the tibial view outside the ambiguous zone or modifications of the model-image registration code to enforce smooth kinematic continuity across image frames and/or to impose joint penetration/separation penalties \cite{muJOINTTRACKOPENSOURCEEASILY2007}. Second, we observed similar measurement performance for the standard and naïve test sets, which differed only in the superior/inferior joint translation. This suggests that the autonomous kinematic processing pipeline can provide reliable measures for implants and imaging systems that were not part of the training set, which will be important for application in novel clinical environments.

Two independent research teams utilized our software to evaluate the accuracy of our autonomous measurement pipeline compared to their reference standard methods using implants and image detectors that were not part of our training sets. In both cases, the accuracy results were comparable to results reported for contemporary human-supervised single-plane model-image registration methods for TKA kinematics \cite{banksAccurateMeasurementThreedimensional1996,floodAutomatedRegistration3D2018, banksVivoKinematicsCruciateretaining1997,banks2003HapPaul2004, komistekVivoFluoroscopicAnalysis2003}. Interestingly, the independent accuracy results appeared superior to our assessment of differences between autonomous and human-supervised measures of TKA kinematics. In both cases, the independent centers used high-resolution flat-panel detectors that provided better spatial resolution and grayscale contrast than most of the imaging systems included in our datasets. With images of similar quality, it is reasonable to expect similar measurement accuracy.  

This work has several limitations. First, the image data sets resulted from previous studies in our labs, so there was no prospective design of which implant systems and image detectors should be included for a pipeline that generalizes well to other implants and detectors. Nevertheless, the naïve data set and the independent assessments, all involving implants and detectors not used for training, performed well and suggest that the method can usefully generalize to measurements of traditionally configured TKA implants. Future work is required to evaluate measurement performance with partial knee arthroplasty or revision implants. Second, many methodologic and configuration options and alternatives remain to be explored, and the current pipeline implementation should not be considered optimal. How best to disambiguate tibial poses and determine the most effective and robust optimization cost functions are areas of current effort.

We present an autonomous pipeline for measuring 3D TKA kinematics from single-plane radiographic images. Measurement reproducibility and accuracy are comparable to contemporary human-supervised methods. We believe capabilities like this will soon make it practical to perform dynamic TKA kinematic analysis in a clinical workflow, where these measures can help surgeons objectively determine the best course of treatment for their patients.

\section{Conflicts of Interest}
None.

\section{Acknowledgements}
This work is supported by a generous donation from the McJunkin Family Charitable Foundation.

\bibliographystyle{IEEEtranN}
\bibliography{bib}

\end{document}